\documentclass[12pt]{iopart}
\usepackage{iopams}

\usepackage{graphicx}
\usepackage{color}
\usepackage{bm}

\newcommand{\eqname}[1]{\label{eq:#1}}

\newcommand{\rr}{ {\bf r}}

\newcommand{\eq}[1]{(\ref{eq:#1})}

\newcommand{\Psihd}{\hat\Psi^\dagger}

\newcommand{\Psih}{\hat\Psi}

\begin{document}

\title[Hawking radiation from acoustic black holes in atomic BECs]
{Numerical observation of Hawking radiation from acoustic black holes in atomic Bose-Einstein condensates 
} 

\author{Iacopo Carusotto $^{1}$, Serena Fagnocchi $^{2,3}$, Alessio Recati $^1$, Roberto Balbinot $^3$, and Alessandro Fabbri $^4$}

\address{$^1$ CNR-INFM BEC Center and Dipartimento di Fisica, 
Universit\`a di Trento,  \\ via Sommarive 14, I-38050 Povo, Trento, Italy}
\address{$^2$ Centro Studi e Ricerche ``Enrico Fermi'', Compendio
Viminale, 00184 Roma, Italy}
\address{$^3$ Dipartimento di Fisica dell'Universit\`a di Bologna and
INFN sezione di Bologna, Via Irnerio 46, 40126 Bologna, Italy}
\address{$^4$ Departamento de Fisica Teorica and IFIC, Universidad
de Valencia-CSIC, C. Dr.Moliner, 50, 46100 Burjassot, Spain}

\ead{carusott@science.unitn.it}

\begin{abstract}
We report numerical evidence of Hawking emission of Bogoliubov phonons from a sonic horizon in a flowing one-dimensional atomic Bose-Einstein condensate. The presence of Hawking radiation is revealed from peculiar long-range patterns in the density-density correlation function of the gas. Quantitative agreement between our fully microscopic calculations and the prediction of analog models is obtained in the hydrodynamic limit. New features are predicted and the robustness of the Hawking signal against a finite temperature discussed.
\end{abstract}

\date{\today}


\section{Introduction}

Back in 1974 S. W. Hawking~\cite{hawking,hawking2} showed that black holes are not completely ``black'' objects, but rather emit thermal radiation at a temperature inversely proportional to their mass. 
This amazing prediction, crucial to establish the connection between black holes and thermodynamics~\cite{bekenstein}, represents a genuine quantum effect in a gravitational context and is widely considered as a milestone of modern theoretical physics. Despite its conceptual importance, the weak intensity of Hawking radiation has so far prevented any direct experimental observation. 

On the basis of a formal analogy between the propagation of waves in inhomogeneous and moving media and the propagation of fields on a curved space-time background, W.G. Unruh predicted in 1981 the occurrence of Hawking radiation in any system developing a horizon for some wavy perturbation~\cite{unruh}.
In the following years, many systems have been proposed as candidates for actual experimental detection of this class of effects~\cite{analogy_book}, e.g. superfluid liquid Helium~\cite{Helium}, atomic Bose-Einstein condensates (BECs)~\cite{bec-analogy-hawk,bec-analogy,wignerBH}, degenerate Fermi gases~\cite{fermi}, slow light in moving media~\cite{slowlight}, travelling refractive index interfaces in nonlinear optical media~\cite{fiber}, surface waves in water tanks~\cite{water}.

In the present paper we shall focus our attention on the specific case of Bogoliubov phonons propagating on top of a moving atomic Bose-Einstein condensate. 
Atomic BECs are among the cleanest system where quantum physics can be investigated~\cite{manybodyBEC,QI,BECbook}: the temperature can in fact be made so low that the behavior of matter is dominated by the dual particle-wave nature of its constituents and the quantum dynamics is not masked by spurious thermal effects. 
Furthermore, in contrast to other quantum coherent condensed-matter systems such as superfluid liquid Helium, quantitative theories able to describe the collective dynamics from a microscopic standpoint are available.

The possibility of creating in these systems black hole-like configurations that allow the study of the analog of Hawking radiation has been discussed in the last years by several authors~\cite{bec-analogy-hawk}.
In a recent work~\cite{bff} based on the gravitational analogy, we predicted that in this case a very characteristic pattern appears in the correlation function for the density fluctuations of a condensate as a consequence of the Hawking effect.
As demonstrated by a number of recent experiments~\cite{jin,HBTlattice,HBT_aspect,other_noise_exp,noise_th,atom_detect_cavity}, the measurement of density correlations appears as a powerful tool to extract information on the microscopic physics of atomic gases.
Since the Hawking effect consists of pairs of correlated phonons being emitted in opposite directions from the horizon, the quantum correlations within a pair propagate across the condensate and result into long-range density correlations between distant points on opposite sides of the horizon.

The present paper reports numerical simulations that nicely confirm our predictions of Ref.\cite{bff} and further support the promise of experimental detection of the Hawking radiation from the density correlations rather than from the phonon flux. We also point out qualitative features that can be used to univocally distinguish the Hawking signal from fluctuations of different nature.
Differently from most previous works on analog models, the present calculations are based on the application of microscopic many-body techniques to an experimentally realistic system. 
Since our numerical calculation never relies on the gravitational analogy, it represents an independent evidence of the existence of the Hawking effect in a realistic condensed-matter system. The quantitative intensity of the Hawking signal is in agreement with the gravitational analogy. 

The structure of the paper is the following.
In Sec.\ref{sec:system} we introduce the physical system under investigation and in Sec.\ref{sec:wigner} we summarize the theoretical method used for the calculations.
The numerical evidence of Hawking radiation is presented in Sec.\ref{sec:Hawk} and quantitatively analyzed in Sec.\ref{sec:quant} where an extensive comparison with analytical results based on the gravitational analogy is made.
The effect of a finite initial temperature on the Hawking effect is discussed in the following Sec.\ref{sec:thermal}. 
A brief discussion of the actual observability of the predicted effect with state-of-the-art systems and detection schemes is given in Sec.\ref{sec:exp}.
Conclusions are finally drawn in Sec.\ref{sec:conclu}.

\section{An acoustic black hole in a flowing condensate}
\label{sec:system}

\begin{figure}[htbp]
\begin{center}
\includegraphics[width=0.6\columnwidth,angle=0,clip]{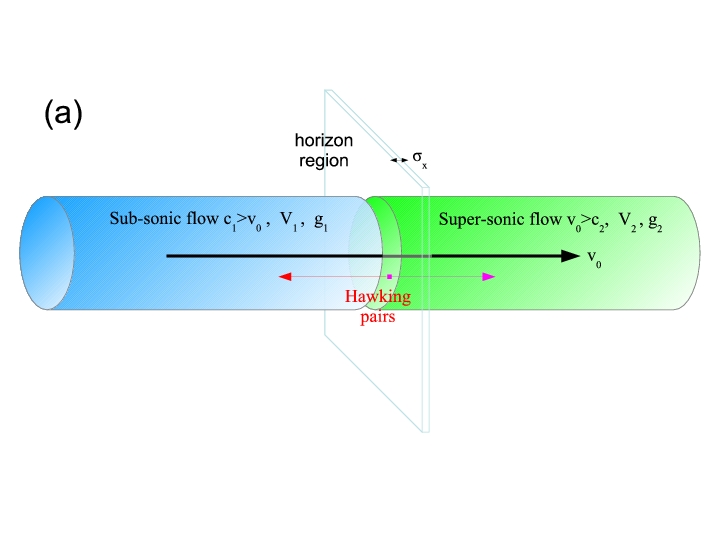}\\
\includegraphics[width=0.8\columnwidth,angle=0,clip]{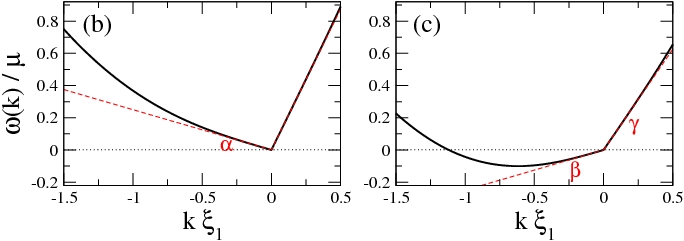}
\caption{
Panel (a): scheme of the physical system under investigation.
Panels (b,c): dispersion of Bogoliubov excitations on top of the flowing
 condensate in the regions of respectively sub- (b) and super-sonic (c)
 flow. System parameters: $v_0/c_1=0.75$, $v_0/c_2=1.5$.
}
\label{fig:disp}
\end{center}
\end{figure}

The physical system that we consider is sketched in Fig.\ref{fig:disp}(a): an elongated atomic Bose-Einstein condensate which is steadily flowing at a speed $v_0$ along an atomic waveguide. The transverse confinement is assumed to be tight enough for the transverse degrees of freedom to be frozen~\cite{BECbook} and the system dynamics to be accurately described by a one-dimensional model based on the following second-quantized Hamiltonian,
\begin{eqnarray}
\mathcal{H}&=&\int\!dx\,\Big[\frac{\hbar^2}{2m}\,\nabla \Psihd(x) \, \nabla\Psih(x)+V(x)\,\Psihd(x)\,\Psih(x)\,+ \nonumber \\ 
&+&\frac{g(x)}{2}\,\Psihd(x)\,\Psihd(x)\,\Psih(x)\,\Psih(x) \Big].
\end{eqnarray}
Here, $\Psih(x)$ and $\Psihd(x)$ are atomic field operators satisfying Bose commutation rules $[\Psih(x),\Psihd(x')]=\delta(x-x')$, $m$ is the atomic mass, $V(x)$ the external potential, and $g(x)$ is the atom-atom interaction constant~\cite{BECbook}.

If the characteristic size $\ell_\perp$ of the transverse wavefunction $\phi(\rr_\perp)$ is much longer than the atom-atom scattering length $a_0$, the interaction constant has the form~\cite{Olshanii,pavloff}:
\begin{equation}
g=\frac{4\pi \hbar^2\,a_0}{m}\,\int\,d^2\rr_\perp\,\left|\phi(\rr_\perp)\right|^4,
\end{equation}
which simplifies to $g=2\,\hbar \omega_\perp\,a_0$ in the most relevant case of a cylindrically symmetric and harmonic transverse trapping potential of frequency $\omega_\perp$ if interactions are weak enough not to distort its Gaussian ground state wavefunction. In this limit, also the contribution to the external potential $V(x)$ due to the zero-point energy of the transverse ground state has a simple form $2\times \hbar \omega_\perp/2$.

Initially, the condensate has a spatially uniform density $n$. The external potential and the (repulsive) atom-atom interaction constant are also uniform and equal to respectively $V(x)=V_1$ and $g(x)=g_1>0$.
Around $t=t_0$, a steplike spatial modulation of characteristic thickness $\sigma_x$ is applied to both the potential and the interaction constant by suitably modifying the transverse confinement potential and/or the atom-atom scattering length via an external magnetic field tuned in the vicinity of a so-called Feshbach resonance~\cite{BECbook}: within a short time $\sigma_t$, $V$ and $g$ in the downstream $x>0$ region are brought to $V_2$ and $g_2$, while their values in the $x<0$ region are kept equal to the initial ones $V_1$ and $g_1$.

In order to suppress competing processes such as back-scattering of condensate atoms and soliton shedding from the potential step~\cite{pavloff}, the external potential $V$ is chosen to exactly compensate the spatial jump in the Hartree interaction energy $\mu_{1,2}=g_{1,2}\,n$, i.e. $V_2+\mu_2=V_1+\mu_1$~\footnote{We have checked that all our conclusions are not sensitive to slight misalignments of $V$ and $g$.}.
In this way, the plane wave
\begin{equation}
\psi(x,t)=\sqrt{n}\,\exp[i(k_0 x-\omega_0 t)]
\eqname{planewave}
\end{equation}
with $\hbar k_0/m =v_0$ and $\omega_0=\hbar k_0^2/2m$ is for all times a solution of the Gross-Pitaevskii equation (GPE)
\begin{equation}
i\hbar\,\frac{\partial \psi}{\partial t}=-\frac{\hbar^2}{2m}\,\frac{\partial^2 \psi}{\partial x^2}+V(x,t)\,\psi+g(x,t)\,|\psi|^2\,\psi.
\eqname{GPE}
\end{equation}
that describes the condensate evolution at the mean-field level~\cite{BECbook}.

In what follows we will focus our attention on the {\em black hole} (or {\em dumb hole}) configuration where the speed of sound $c_{1,2} =\sqrt{\mu_{1,2}/m}$ in the different regions satisfies the chain inequality $c_1 >v_0 >c_2$: a black hole-type sonic horizon separates a region of subsonic $c_1>v_0$ flow outside the black hole from a supersonic $v_0>c_2$ one inside the black hole. 
As one can see from the Bogoliubov dispersion plotted in Fig.\ref{fig:disp}(c), long-wavelength phonon excitations propagating in the supersonic region are dragged away by the moving condensate and are then unable to propagate back to the horizon. Only higher-$k$, single-particle excitations outside the hydrodynamical window can emerge. 

In agreement with previous work~\cite{barcelo}, dynamical stability of this black hole configuration has been numerically verified. A more detailed discussion on stability issues is postponed to a forthcoming publication~\cite{nicolas}.

\section{The Wigner method}
\label{sec:wigner}

The dynamics of fluctuations around the plane-wave mean field solution \eq{planewave} during and after the formation of the horizon can be numerically studied by means of the so-called truncated Wigner method for the interacting Bose field~\cite{QO,wigner0}.
Application of this technique to calculate the time-evolution of generic observables of an interacting Bose gas has been extensively developed and validated in~\cite{Yvan_wigner}. For dilute gases such that $n\xi^d\gg 1$ (the healing length is defined as usual as $\xi=\sqrt{\hbar^2/m\mu}$ and $d$ is the dimensionality of the system, $d=1$ in our case), this technique is equivalent to the time-dependent Bogoliubov approach and has the advantage of being able to follow the system for longer times when the backaction of quantum fluctuations on the condensate starts to be important.
A first application of this method to atomic BEC-based analog models is reported in~\cite{wignerBH}.

At $t=0$ well before the formation of the black-hole horizon, the condensate is assumed to be uniform and at thermal equilibrium in the moving frame at $v_0$ at a temperature $T_0$. 
Within the Wigner framework, this corresponds to taking a random initial wavefunction $\psi_0(x)$ of the form: 
\begin{equation}
\psi_0(x)=e^{i(k_0 x-\omega_0 t)}\, \{ \sqrt{n_0}+ \sum_{k\neq0} [\alpha_k\,u_k\,e^{ikx}+ \alpha_k^*\,v_k\,e^{-ikx}].
\},
\end{equation}
As both the external potential and the interaction constant are spatially uniform $V(x)=V_1$ and $g(x)=g_1$, the Bogoliubov modes have the standard plane-wave form. The Bogoliubov coefficients $u_k,v_k$ are defined in terms of the kinetic and Bogoliubov energies $E_k=\hbar^2 k^2/2m$ and
$\epsilon_k=\sqrt{E_k(E_k+2g_1 n)}$ by $u_k\pm v_k=(E_k/\epsilon_k)^{\pm 1/4}$. 

The mode amplitudes $\alpha_k$ are independent, zero-mean Gaussian random variables such that $\langle \alpha_k \rangle= \langle \alpha_k^2 \rangle=0$. The variance $\langle |\alpha_k|^2 \rangle=[2\tanh(\epsilon_k/2\,k_B\,T_0)]^{-1}$ tends to a finite value $1/2$ in the $T_0\rightarrow 0$ limit, so to include zero-point fluctuations. For each realization of the random amplitudes $\alpha_k$'s, the condensate density $n_0$ has to be suitably renormalized so to account for the condensate depletion~\cite{Yvan_wigner}.

The random classical wavefunction $\psi(x,t)$ is then propagated in time according to the standard GPE \eq{GPE} starting from its initial value $\psi(x,t=0)=\psi_0(x)$ and including the full time- and space-dependence of $V(x,t)$ and $g(x,t)$.
Periodic boundary conditions are assumed for the spatial variable $x$, but an absorbing region far from the horizon is required to prevent the onset of spurious dynamical instabilities due to excitations circulating around the integration box~\cite{barcelo,hawklaser}.
Expectation values of symmetrically-ordered field observables at any later time $t$ are finally obtained as the corresponding averages of the classical field $\psi(x,t)$ over the random initial condition $\psi_0$(x). As usual, trivial one-time commutators have to be subtracted out if normally-ordered quantities are required.

\section{Numerical evidence of Hawking radiation}
\label{sec:Hawk}

\begin{figure}[htbp]
\begin{center}
\includegraphics[width=0.5\columnwidth,angle=0,clip]{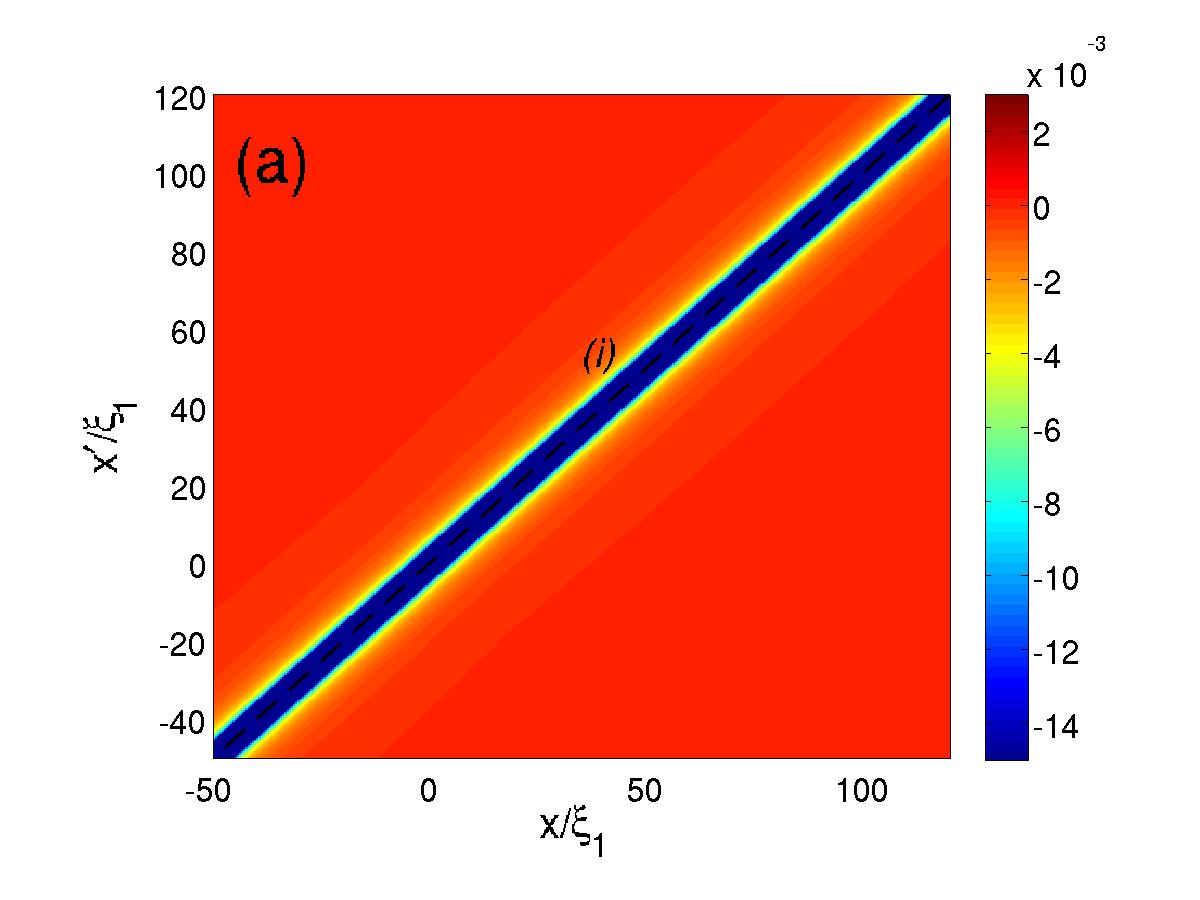}
\includegraphics[width=0.5\columnwidth,angle=0,clip]{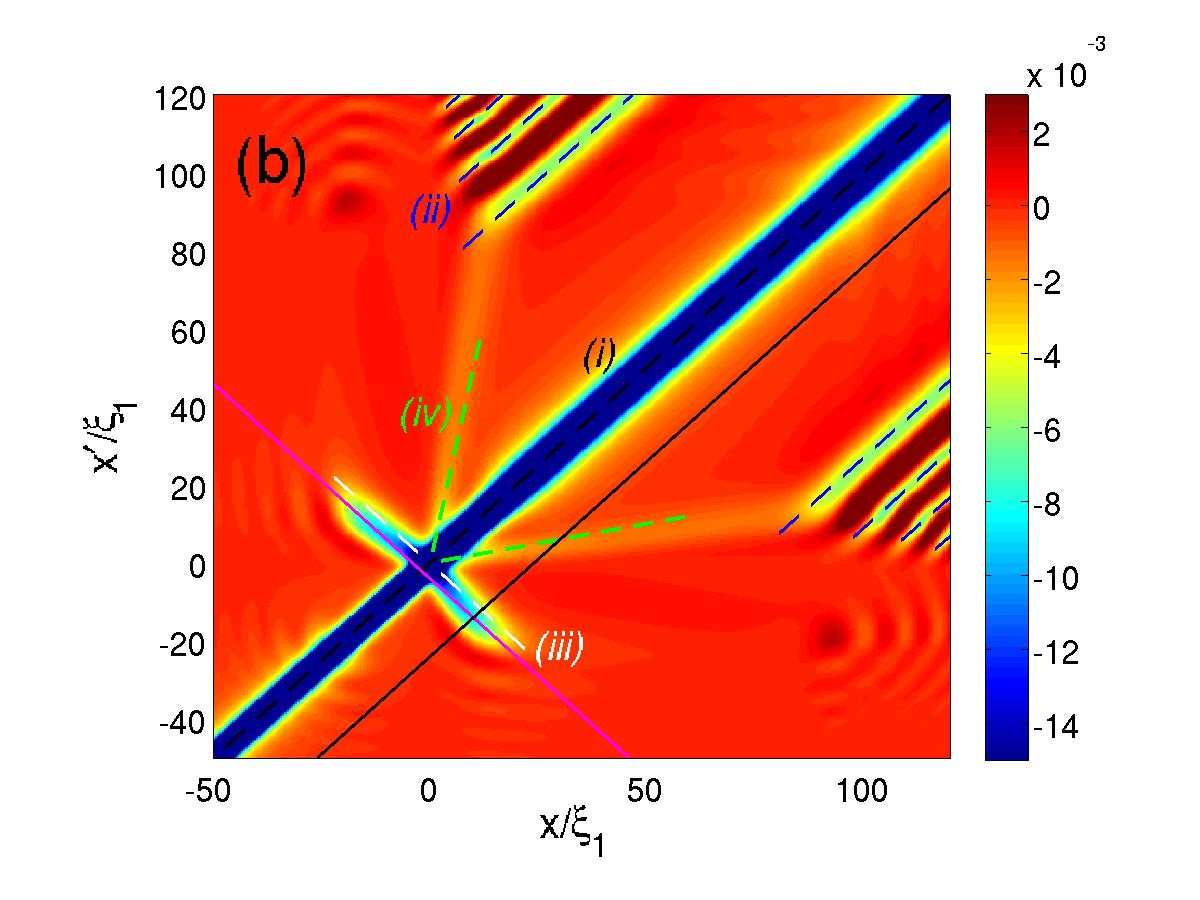}
\includegraphics[width=0.5\columnwidth,angle=0,clip]{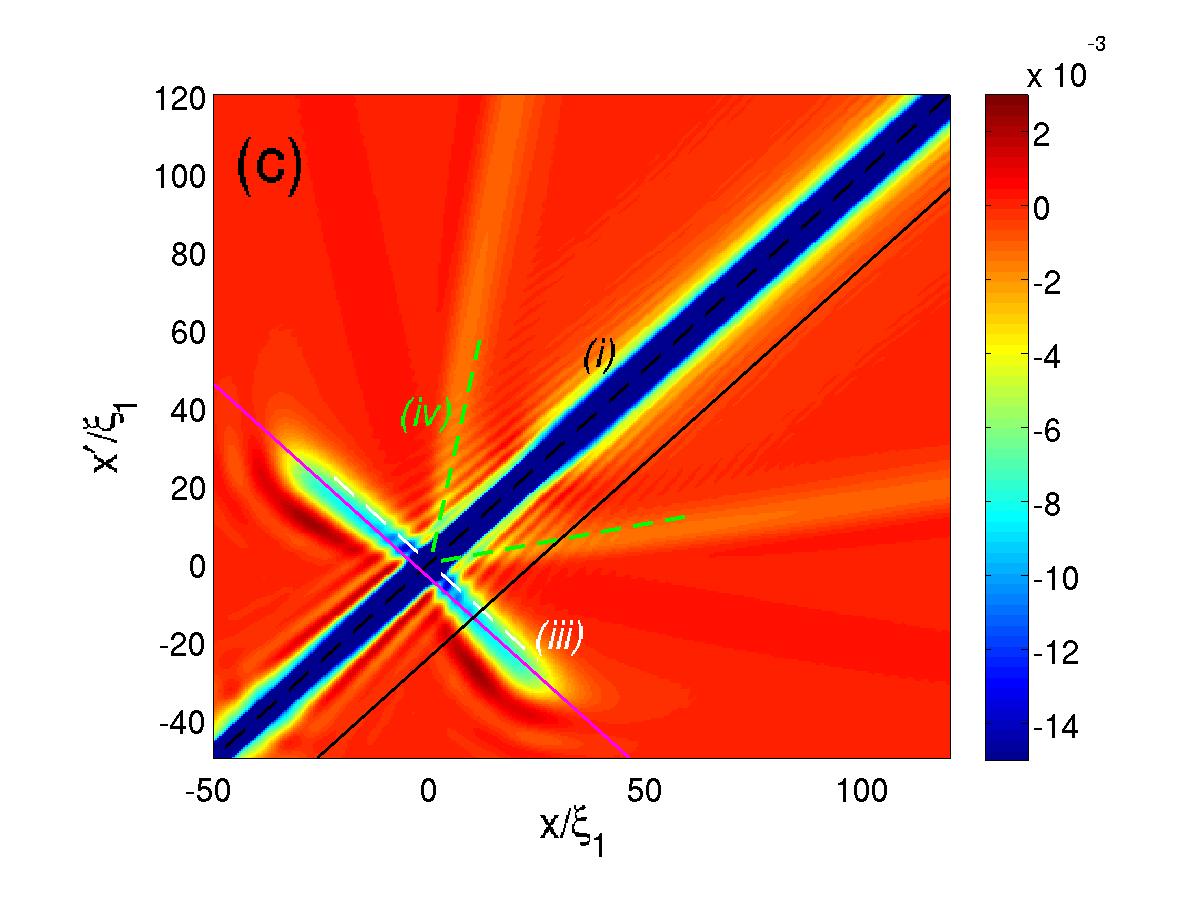}
\end{center}
\caption{
Density plot of the rescaled density correlation $(n\xi_1)\times[G^{(2)}(x,x')-1]$ at the initial time $\mu_1 t = 0$ (a) and at two successive times $\mu_1 t=70, 120$ well after the switch-on of the horizon (b,c).
The dashed lines and the $(i)$, $(ii)$, $(iii)$, $(iv)$ labels identify the main features discussed in the text. 
The solid black and magenta lines in panels (b,c) indicate the directions along which the cuts shown in Fig.\ref{fig:g2_cut} are taken.
The horizon is formed within a time $\mu_1 \,\sigma_t=0.5$ around $\mu_1\,t_0=2$ and has a spatial width $\sigma_x/\xi_1=0.5$. System parameters as in Fig.\ref{fig:disp}. The initial temperature is $T_0=0$. 
}
\label{fig:g2}
\end{figure}

\begin{figure}[htbp]
\begin{center}
\includegraphics[width=0.5\columnwidth,angle=0,clip]{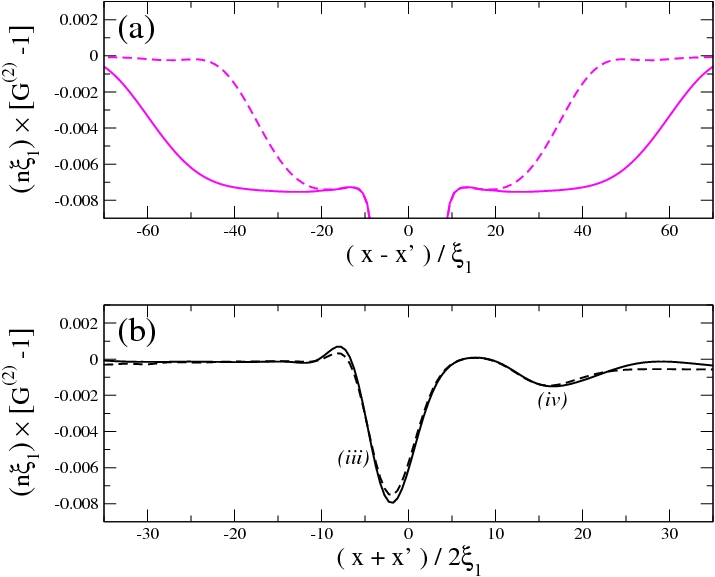}
\end{center}
\caption{
Cuts of the reduced density correlation at times $\mu_1 t=70$ (dashed lines) and $120$ (solid lines) taken along the magenta ($x+x'=-5.5\xi_1$) and black ($x=x'+23.5\,\xi_1$) lines of Fig.\ref{fig:g2}(b,c).
}
\label{fig:g2_cut}
\end{figure}

Our numerical study of the Hawking radiation from the acoustic black hole is based on a vast campaign of Wigner simulations of the condensate evolution after the formation of the black-hole horizon.
Inspired by our recent work~\cite{bff}, specific attention has been paid to the correlation function of density fluctuations, i.e. the normalized, normal-ordered density-density correlation function:
\begin{equation}
G^{(2)}(x,x')=\frac{\langle : n(x)\,n(x'): \rangle}{\langle n(x)\rangle\,
\langle n(x')\rangle}.
\eqname{g2_def}
\end{equation}
Three successive snapshots of $G^{(2)}(x,x')$ before (a) and after (b,c) the horizon formation are shown in Fig.\ref{fig:g2}. 
A complete movie of the time-evolution is available online as Supplementary Information~\footnote{To improve the quality of the figures and suppress numerical artefacts, a Gaussian spatial averaging has been performed on the numerical data with an averaging length $\ell_{av}=2$. We have checked that this does not introduce any further artefact except a small quantitative reduction of the peak signal intensity.}
 
The main features that are observed in the figure can be classified as follows:

\begin{itemize}

\item[$(i)$] A strong, negative correlation strip is always present along the main diagonal $x=x'$ and is almost unaffected by the horizon formation process.

\item[$(ii)$] A system of fringes parallel to the main diagonal appears inside the black hole as soon as the horizon is formed. As time goes on, these fringes move away from the main diagonal at an approximately constant speed and eventually disapper from the region of sight. 

\item[$(iii)$] Symmetric pairs of negative correlation tongues extend from the horizon point almost orthogonally to the main diagonal. While their maximum height remains almost constant in time, their length linearly grows with time [see also Fig.\ref{fig:g2_cut}(a,b)]. These tongues involve pairs of points located on opposite sides of the horizon.

\item[$(iv)$] A second pair of tongues appears for pairs of points located inside the black hole. Both their height and length scale in the same way as for feature $(iii)$. 

\end{itemize}

\begin{figure}[htbp]
\begin{center}
\includegraphics[width=0.5\columnwidth,angle=0,clip]{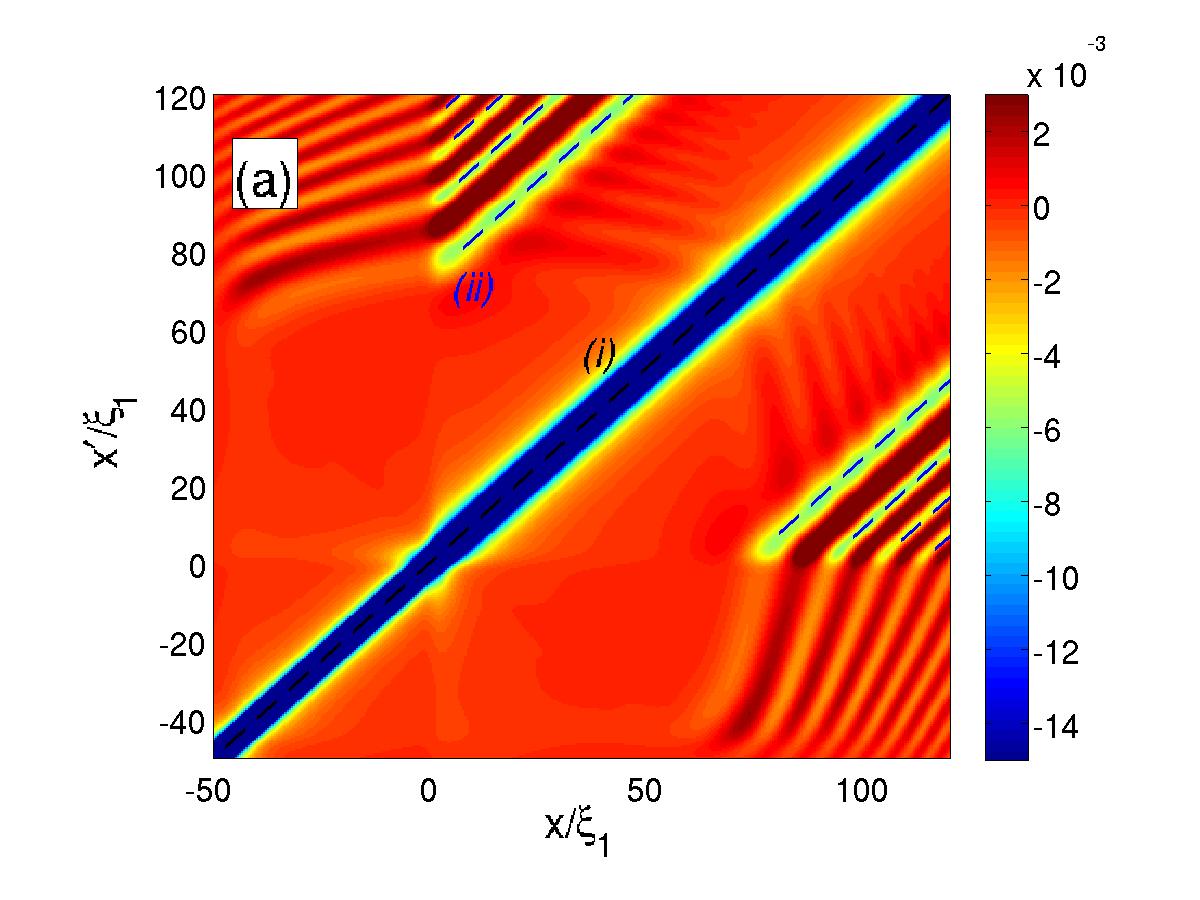}
\includegraphics[width=0.5\columnwidth,angle=0,clip]{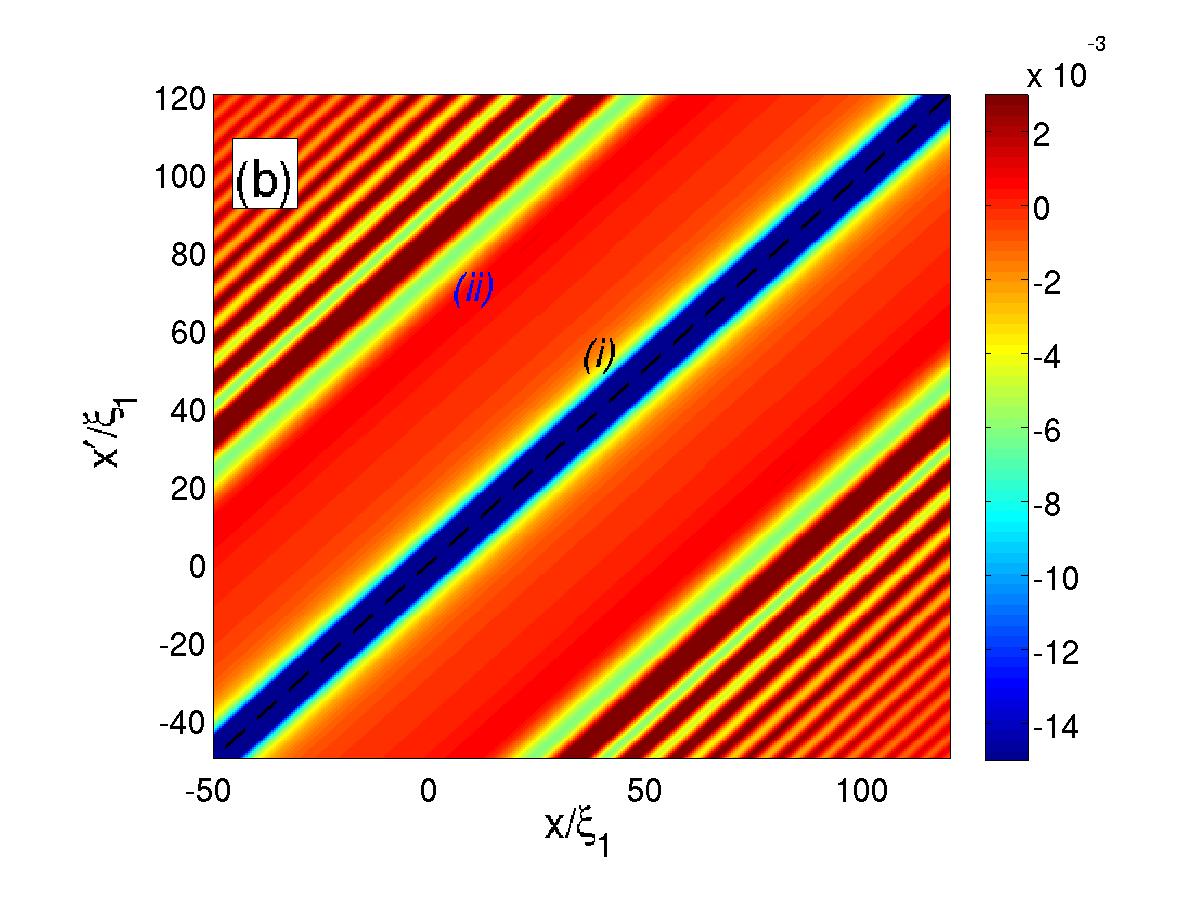}
\end{center}
\caption{
Panel (a): density plot of the rescaled density correlation \mbox{$(n\xi_1)\times[G^{(2)}(x,x')-1]$} in the absence of the black hole horizon, the flow being everywhere sub-sonic $v_0/c_1=0.4$, $v_0/c_2=0.8$. 
Panel (b): the same quantity for a spatially homogeneous system whose interaction constant $g$ is varied in time from $g_1$ to $g_2$ with the same functional law as in Fig.\ref{fig:g2}.
}  
\label{fig:casimir_subsonic}
\end{figure}

Apart from feature $(i)$ which is the usual antibunching due to the repulsive atom-atom interactions~\cite{antibunching}, these observations illustrate a variety of effects of quantum field theory in a spatially and temporally varying background~\cite{c-QFT}.
 
Feature $(ii)$ is a transient effect that originates in the bulk of the internal $x>0$ region and has no relation with the presence of a horizon.
An identical fringe pattern (yet extending to the whole system, rather than to the $x>0$ region only) is in fact obtained if a spatially uniform system is considered whose interaction constant is varied in time from $g_1$ to $g_2$ with the same functional law [Fig.\ref{fig:casimir_subsonic}(b)].
Its physical interpretation is the following: as a consequence of the time-modulation of the interaction constant $g$, correlated pairs of Bogoliubov phonons are generated during the short modulation time by a phonon analog of the dynamical Casimir effect~\cite{casimir,CasimirBEC_exp,CasimirBEC,CasimirBEC2}, a quantum process which shares many analogies with the parametric emission of Faraday waves in classical fluids~\cite{faraday}.

As the emission process takes place in a simultaneous and coherent way at all spatial positions, the two phonons are emitted with opposite wavevectors and then propagate in opposite directions. Their quantum correlations reflect into a pattern that only depends on the relative coordinate $x-x'$ and consists of fringes parallel to the $x=x'$ line which move away in time at approximately twice the speed of sound. 
As usual, the longer the ramp time $\sigma_t$, the weaker this dynamical Casimir signal which eventually disappears in the limit of a very long $\sigma_t$.

On the other hand, features $(iii)$ and $(iv)$ do not depend on $\sigma_t$, but only on the eventual presence of a horizon at long times. In particular, they completely disappear if the flow remains everywhere sub-sonic $v_0<c_{1,2}$ [Fig.\ref{fig:casimir_subsonic}(a)] or, {\em a fortiori}, if a spatially uniform system is considered [Fig.\ref{fig:casimir_subsonic}(b)].
This fact, together with their shape in the $(x,x')$ plane and their persistance for indefinite times after the horizon formation suggests a strict link with the Hawking effect.

As anticipated in~\cite{bff}, the quantum correlations within a pair of Hawking phonons emitted in respectively the inward and outward direction translate into a correlation between the density fluctuations at distant points located on opposite sides from the horizon.
Once the horizon is formed, correlated pairs of Hawking phonons are continuously emitted at all times $t>t_0$ on the $\alpha$ and $\beta$ phonon branches of Fig.\ref{fig:disp}(b,c). These phonons then propagate from the horizon in respectively the outward and inward direction at speeds $v_0-c_1<0$ and $v_0-c_2>0$.
A generic time $\tau$ after their emission they are located at $x=(v_0-c_1)\,\tau<0$ and $x'=(v_0-c_2)\,\tau > 0$, which defines a straight line of slope 
\footnote{As the density correlation patterns is symmetric under the exchange $x\leftrightarrow x'$, every tongue in the $x'>x$ half-plane above the main diagonal corresponds to a symmetric one in the half-plane $x'<x$ below the main diagonal. When we speak about the slope of a tongue, we implicitly refer to the upper one in the $x'>x$ region.}
$(v_0-c_1)/(v_0-c_2)$.
As one can see in Fig.\ref{fig:g2}(b), this line (indicated as a white dashed line) almost coincides with the axis of the numerically observed tongue $(iii)$.

Feature $(iv)$ originates from the (partial) elastic back-scattering of the $\alpha$ Hawking partner onto branch $\gamma$: both $\gamma$ and $\beta$ phonons eventually propagate in the inward direction  at speeds $v_0\pm c_2$. Again, the analytically predicted slope $(v_0-c_2)/(v_0+c_2)$ (green dashed lines) well agrees with the axis of the numerically observed tongue $(iv)$. 

\section{Quantitative analysis}
\label{sec:quant}

\begin{figure}[htbp]
\begin{center}
\includegraphics[width=0.5\columnwidth,angle=0,clip]{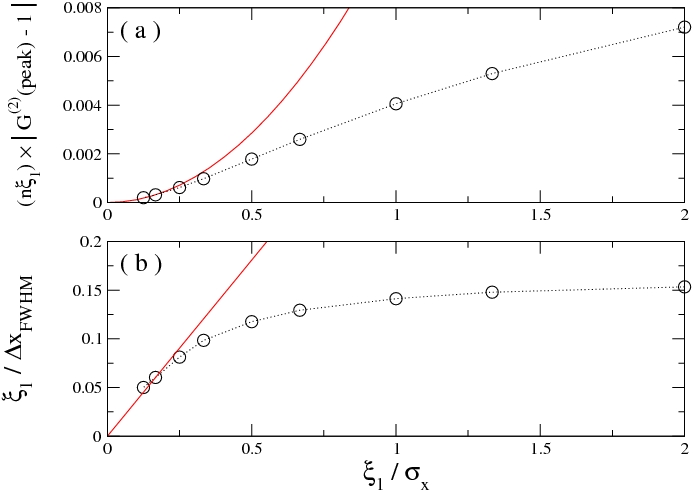}
\end{center}
\caption{
Normalized peak intensity (a) and inverse FWHM (b) of the
 Hawking tongue $(iii)$ as a function of the 
surface gravity of the horizon.
Circles: numerical results. Solid lines: prediction of the gravitational
 analogy. Numerical uncertainty is of the order of the symbol size.
Other system parameters as in Fig.\ref{fig:g2} and Fig.\ref{fig:g2_cut}.} 
\label{fig:dependences}
\end{figure}

The identification of features $(iii)$ and $(iv)$ as signatures of Hawking radiation is further confirmed by a quantitative comparison of our numerical data with the predictions~\cite{bff} of the gravitational analogy.
The peak value $G^{(2)}(\textrm{peak})$ and the transverse full width at half maximum (FWHM) $\Delta x_{FWHM}$ of the tongue $(iii)$ have been extracted from the cut of $G^{(2)}(x,x')$ taken along a straight line $x=x'+x_{cut}$ well outside the antibunching dip $(i)$ [indicated as a black line in Fig.\ref{fig:g2}(b,c)]. 
Examples of such cuts are shown in Fig.\ref{fig:g2_cut}(b). 
Once the tip of the tongue has crossed the cut line, neither the peak value nor the FWHM depend any longer on the position $x_{cut}$ of the cut line nor on the time $t$ of the observation: as expected, the Hawking emission is in fact stationary in both space and time.

In Fig.\ref{fig:dependences}(a), the peak intensity of the correlation signal $(iii)$ is plotted as a function of the inverse of the thickness $\sigma_x$ of the horizon region, a quantity that in the gravitational analogy is proportional to the so-called surface gravity.
As expected, the agreement with the gravitational prediction~\cite{bff}:
\begin{equation}
G^{(2)}(\textrm{peak})= 1-\frac{\kappa^2\,\xi_1 \xi_2}{16\pi\, c_1
 c_2}\,\frac{1}{\sqrt{(n \xi_1)(n \xi_2)}} \frac{c_1\,c_2}{(c_1-v_0)(v_0-c_2)} 
\end{equation}
is quantitatively excellent in the hydrodynamic limit $\sigma_x/\xi_{1,2} \gg 1$ where the physics of the many-body system is dominated by the hydrodynamic modes that are included in the gravitational analogy. 
As usual, the parameter $\kappa$ proportional to the surface gravity of the horizon is defined as
\begin{equation}
\kappa=-\frac{1}{2v}\frac{d}{dx}(c^2-v^2)|_H.
\end{equation}
In the plots, we have shown the renormalized quantity $(n\xi_1)\times [G^{(2)}(x,x')-1]$ that is universal in the dilute gas limit $n\xi_1\gg 1$ where our Wigner approach is exact. The actual intensity of the Hawking signal is then inversely proportional to the dilution parameter $n\xi_1$. Although a more sophisticate theoretical approach may be required to obtain quantitative predictions in the strongly interacting $n\,\xi_1\gtrsim 1$ case, the qualitative features of Hawking physics are not expected to change if a relatively strongly interacting system is used in order to maximize the intensity of the Hawking signal. 

The inverse FWHM of the tongue $(iii)$ in the transverse direction is plotted in Fig.\ref{fig:dependences}(b) as a function of the inverse thickness of the horizon region. Again, the agreement of the numerical result with the gravitational prediction of~\cite{bff} is very good as long as $\sigma_x/\xi_1\gg 1$.
This observation appears even more significant if one remembers that in this framework the FWHM is a good measure of the surface gravity $\kappa$, which then fully determines the Hawking temperature $T_H=\hbar \kappa /(2\pi k_B)$~\cite{c-QFT}.

\section{Effect of a finite initial temperature}
\label{sec:thermal}

\begin{figure}[htbp]
\begin{center}
\includegraphics[width=0.5\columnwidth,angle=0,clip]{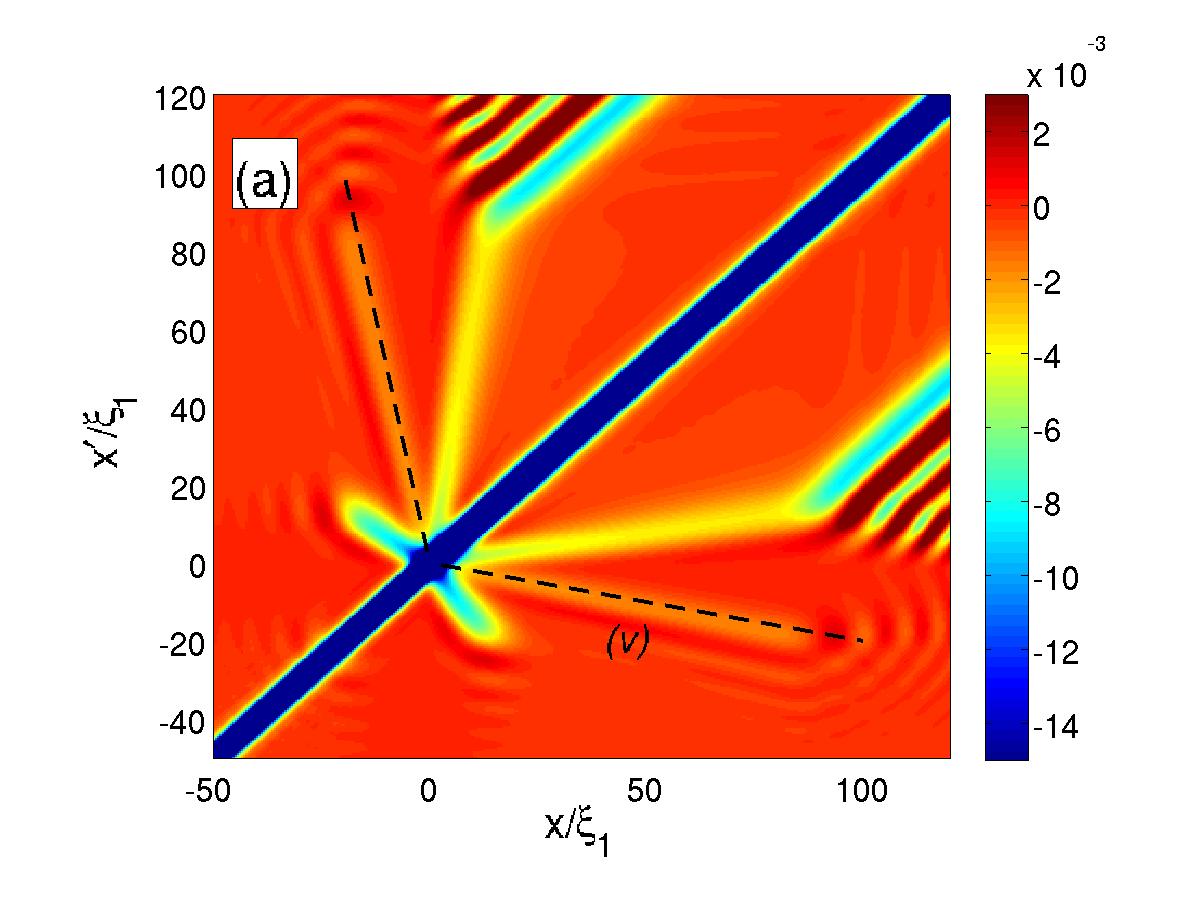}
\includegraphics[width=0.5\columnwidth,angle=0,clip]{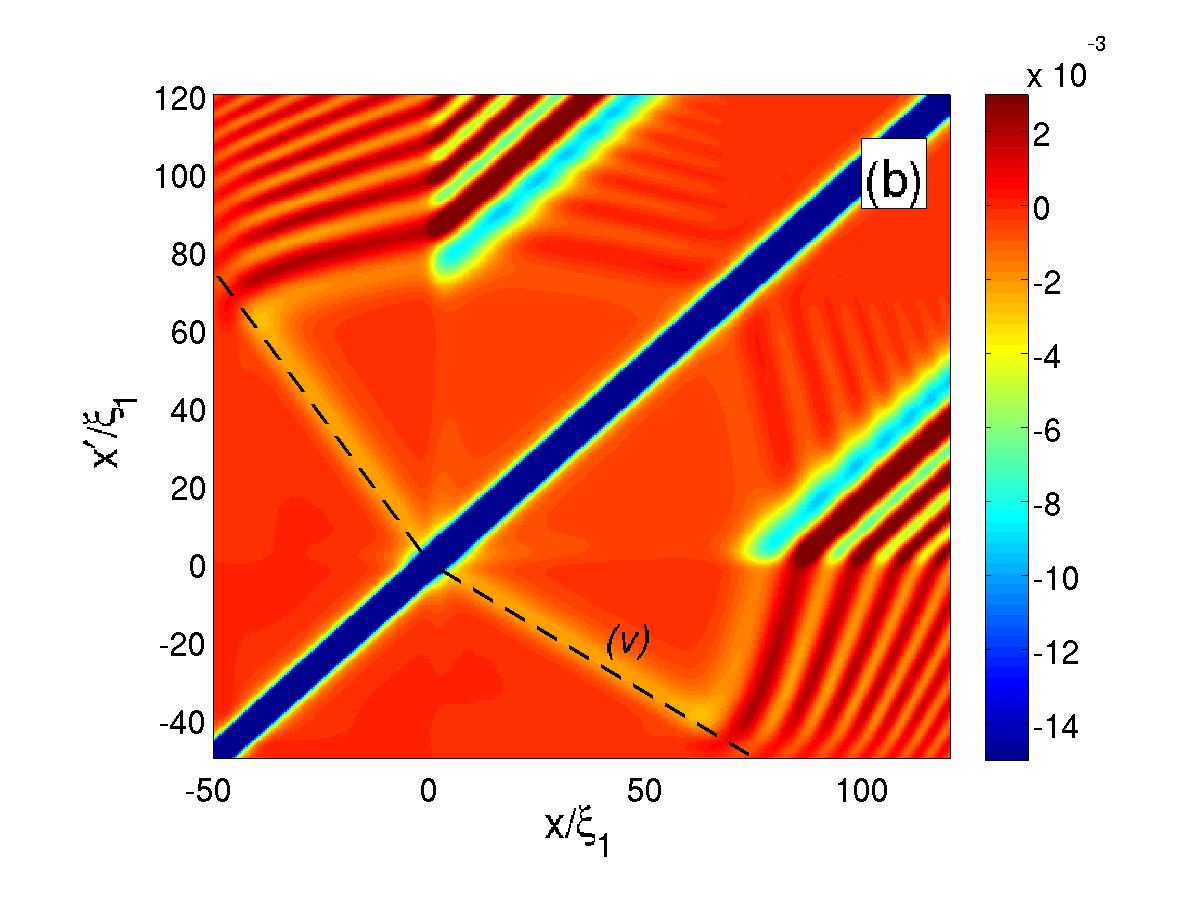}
\end{center}
\caption{
Panel (a): density plot of the rescaled density correlation \mbox{$(n\xi_1)\times[G^{(2)}(x,x')-1]$} for the same parameters as in Fig.\ref{fig:g2}(b) but a finite initial temperature $k_B T_0/\mu_1=0.1$.
Panel (b): the same quantity in the absence of the black hole horizon, the flow being everywhere sub-sonic $v_0/c_1=0.4$, $v_0/c_2=0.8$ as in Fig.\ref{fig:casimir_subsonic}(a). Initial temperature $k_B T_0/\mu_1=0.1$.}  
\label{fig:finiteT}
\end{figure}

To verify the robustness of our observations with respect to thermal effects, we have performed a series of numerical calculations starting from a finite initial temperature $T_0>0$. The results are summarized in Fig.\ref{fig:finiteT}.

As it happens for thermal light at semi-reflecting mirrors~\cite{QO}, the partial reflection of thermal phonons on the horizon provides a further mechanism for the appearance of non-trivial density correlation between the inner $x>0$ and the outer $x<0$ regions which may mask the Hawking signal.
This concern is ruled out by the numerical results shown in Fig.\ref{fig:finiteT}(a). 
Even for an initial temperature $k_B T /\mu_1=0.1$  higher than the Hawking temperature $k_B T_H/\mu_1 \simeq 0.05$ expected from the gravitational analogy~\cite{analogy_book}, the $(iii)$ and $(iv)$ Hawking tongues remain perfectly visible and are even a bit strengthened by stimulation effects~\cite{beken_Tf}.
The new feature $(v)$ that originates from thermal effects is located between the tongues $(iii)$ and $(iv)$ and is well distinct from them. Its attribution to partial reflection of thermal phonons is confirmed by the value of its slope, which is in very good agreement with the analytical prediction $(v_0-c_1)/(v_0+c_2)$ indicated as a dashed black line in Fig.\ref{fig:finiteT}(a,b). 

As the slope of the thermal tongue is completely different from the one $(v_0-c_1)/(v_0-c_2)$ of the Hawking tongue, the Hawking signal can be easily isolated in a correlation image from spurious thermal and atom loss~\cite{3body} effects also at temperatures higher than $T_H$. This is even more remarkable as in this regime it appears difficult to separate the Hawking phonons from the thermal ones by looking just at the total phonon flux.

As a final check we have performed a numerical calculation for the case of a flow which remains everywhere sub-sonic. The result is shown in Fig.\ref{fig:finiteT}(b): as expected, the Hawking tongues disappear, while the dynamical Casimir fringes $(ii)$ and the thermal tongue $(v)$ persist without being dramatically affected.

\section{Discussion of some experimental issues}
\label{sec:exp}

We conclude with a brief discussion of the main issues that may arise in an actual experiment that aims to observe the Hawking radiation from the density correlations.

As the magnitude of the Hawking signal is quite low, of the order of $(n\xi_1)\times G^{(2)}(\textrm{peak})\approx 5\times 10^{-3}$, a critical point in an actual experiment will be the signal to noise ratio in extracting the correlation signal due to the Hawking effect from experimental noise. While many sources of noise can be avoided by carefully engineering the experimental set-up, shot noise is an intrinsic consequence of the discrete (i.e. quantum) nature of atoms and therefore can not be eliminated.

A precise measurement then requires a large statistics to be collected, either by repeating the experiment many times or by creating a large number of (almost) identical BEC samples on which to simultaneously perform the measurement. Most probably, both these strategies will face the difficulty of keeping the parameters of the experimental set-up constant in space and/or time.

Another possibility is to use a continuous-wave atom laser beam propagating along an atomic fiber~\cite{atomlaser} and then perform a time average of the correlation signal over long times. Several different outcoupling mechanism have been demonstrated, which allows for a wide tuning of both the density and the flow speed of the atom laser beam. Furthermore, active stabilization techniques~\cite{stabilization} can be used to keep the parameters of the output beam stable in time during the experiment.

In this perspective, the key experimental problem that remains so far unsolved is the one of continuously reloading the mother BEC while emitting the atom laser beam. Existing devices~\cite{atomlaser,atomlaser2} are in fact limited by the finite number of atoms present in the mother BEC, presently of the order of $10^5$, but attempts to continuously reload the condensate during the atom laser operation have been recently reported~\cite{cw_at_laser}.

As the Hawking signal is inversely proportional to the diluteness parameter $n\xi_1$, a relatively strongly interacting system may be favourable to maximize the signal to noise ratio. As our predictions are basically a consequence of quantum hydrodynamics, we expect that their validity extends well beyond the case of weakly interacting BECs that has been considered by our microscopic model. Although no complete theory is available for a strongly interacting gas, the theory of Luttinger liquids~\cite{density_phase} appears to indicate that larger correlation signals can be observed in stronger interacting systems where the diluteness parameter is smaller.

Another crucial issue is to use an atomic detection scheme able to efficiently measure the density correlations from which to extract the Hawking signal: several techniques have been developed during the last years to experimentally characterize local density fluctuations in ultracold atomic gases based on noise in absorption images~\cite{jin,HBTlattice}, on microchannel plate detectors~\cite{HBT_aspect}, or even on high-finesse optical cavities~\cite{atom_detect_cavity}.
Thanks to its non-destructive nature, this last technique to detect single atoms in an atom laser beam appears as a very promising one for the detection of Hawking radiation from density correlations.

\section{Conclusions}
\label{sec:conclu}

In conclusion, we have provided clear numerical evidence of the presence of Hawking radiation in a flowing atomic Bose-Einstein condensate in an acoustic black hole configuration. The signature of this effect was identified in the density-density correlation function for distant points located on opposite sides of the horizon. The intensity of the observed signal is in quantitative agreement with the predictions of the gravitational analogy in the hydrodynamic regime.

A crucial novelty of this work consists in the fact that all numerical results have been obtained starting from a microscopic description of the system dynamics using standard methods of many-body theory, and the gravitational analogy has provided a physical insight to interpret the numerical observations.
In this way, our numerical observations can be considered as an independent proof of the existence of Hawking radiation and rule out some frequent concerns~\cite{high-k} on the role of short wavelength, ``trans-Planckian'' physics on the Hawking emission.

We have shown that the stationary emission of phonons from the acoustic horizon persists even outside the hydrodynamical regime when the gravitational analogy can no longer be applied, and shares many of the typical properties of the Hawking radiation.

The characteristic pattern of the Hawking signal in the density correlation function allows an easy identification from spurious effects of, e.g., thermal origin even at temperatures significantly higher than the Hawking temperature.
This supports our proposal of using density correlations as a powerful experimental tool to detect Hawking radiation from acoustic black holes in atomic Bose-Einstein condensates in the next future.

\section*{Acknowledgements}

We are indebted to Y. Castin, C. Ciuti, R. Floreanini, C. Lobo, Z. Gaburro, N. Pavloff, R. Parentani, M. Rinaldi, and W. Unruh for stimulating discussions.
IC and AR acknowledge support from the italian MIUR and the EuroQUAM-FerMix program. 
IC acknowledges support from the french CNRS. 
SF acknowledges support from E. Fermi center. 
AF acknowledges the spanish MEC.

\appendix

\end{document}